\documentclass[pdflatex,sn-mathphys-num,iicol]{sn-jnl}

\usepackage{graphicx}%
\usepackage{multirow}%
\usepackage{amsmath,amssymb,amsfonts}%
\usepackage{amsthm}%
\usepackage{mathrsfs}%
\usepackage[title]{appendix}%
\usepackage{xcolor}%
\usepackage{textcomp}%
\usepackage{manyfoot}%
\usepackage{booktabs}%
\usepackage{algorithm}%
\usepackage{algorithmicx}%
\usepackage{algpseudocode}%
\usepackage{listings}%

\usepackage{tabularx} 
\usepackage{enumitem}

\theoremstyle{thmstyleone}%
\theoremstyle{thmstyletwo}%

\theoremstyle{thmstylethree}%

\begin{document}

\title[]{Identification of 2{D} colloidal assemblies in images: a threshold processing method versus machine learning}


\author[1]{\fnm{Ludia T.} \sur{Khusainova}}

\author*[1]{\fnm{Konstantin S.} \sur{Kolegov}}\email{k.kolegov87@gmail.com}


\affil*[1]{\orgdiv{Mathematical Modeling Lab}, \orgname{Astrakhan Tatishchev State University}, \orgaddress{\street{20a~Tatishchev~str.}, \city{Astrakhan City}, \state{Astrakhan Region}, \country{Russia}, \postcode{414056}}}


\abstract{This paper is devoted to the problem of identification of colloidal assemblies using the example of two-dimensional coatings (monolayer assemblies). Colloidal systems are used in various fields of science and technology, for example, in applications for photonics and functional coatings. The physical properties depend on the morphology of the structure of the colloidal assemblies. Therefore, effective identification of particle assemblies is of interest. The following classification is considered here: isolated particles, dimers, chains and clusters. We have studied and compared two identification methods: image threshold analysis using the OpenCV library and machine learning using the YOLOv8 model as an example. The features and current results of training a neural network model on a dataset specially prepared for this work are described. A comparative characteristic of both methods is given. The best result was shown by the machine learning method (97\% accuracy).  The threshold processing method showed an accuracy of about 67\%. The developed algorithms and software modules may be useful to scientists and engineers working in the field of materials science in the future.}

\maketitle

\section{Introduction}

There are numerous image processing methods related to particle analysis. One large group of methods is associated with thresholding (binarization)~\cite{Otsu1979,Sankur2004}. The main difference between these methods lies in the algorithms for determining the threshold value. Another group of methods is associated with the use of machine learning and neural networks~\cite{Hannel2018,Cichos2020,Altman2020,Shao2020,Turkmen2023}. Particle detection in images, the study of particle clusters, and the processing of such data are carried out in various fields and applications: analysis of protein structure in medicine~\cite{AlAzzawi2019}, plasma studies in physics~\cite{Mohr2019}, automatic cleaning of a processed part's surface in materials science~\cite{Long2022}, automation of salt precipitate analysis in biological applications~\cite{Batista2025}, analysis of colloidal structures in nanotechnology~\cite{Long2014,Carstensen2018,Newby2018,Boattini2019}, and so on. Processing images of patterns formed by dried droplets of biofluids allows for the diagnosis of various diseases~\cite{Pal2019} and the analysis of dairy product quality~\cite{MolinaCourtois2025}. Image analysis is complicated by the fact that particles can have different shapes, such as spheres, cuboids, or rods, or their mixtures~\cite{LpezGutirrez2022,Fukuda2023}. The authors of~\cite{Pal2024} used image processing and a variety of statistical parameters to study the patterns of dried droplets.

The self-assembly of colloidal structures has become an important manufacturing method for nanostructured materials with a wide range of applications, from optoelectronics to the detection of chemical and biological analytes or the creation of biomimetic surfaces with specific properties or wettability characteristics~\cite{Lotito2017}. Colloidal assemblies, due to their unique nature, are the focus of intense attention in the scientific and engineering community. This relevance is driven not only by their broad spectrum of applications in various fields but also by the potential of these systems to address important challenges of the modern world.

One of the challenges facing researchers is the efficient and accurate identification of colloidal assemblies. Precise segmentation and classification of these assemblies are key steps for understanding their properties and their impact on the environment. Colloidal assemblies, which are dispersed systems with particles ranging in size from a hundred nanometers to several tens of micrometers, are attracting increasing research interest due to their unique properties and wide range of applications. However, issues related to the identification and classification of colloidal assemblies remain relevant and require attention. The aim of this work is to compare two methods for identifying colloidal structures: 1) thresholding in image processing and 2) machine learning.

\section{Methods}

\subsection{Physical problem statement}
The morphology of colloidal particle deposits formed from evaporating droplets and films determines the physical properties of such structures~\cite{Bacchin2018,Zang2019,Kolegov2020}. This refers to the optical~\cite{Inoue2020}, thermal~\cite{Kameya2023}, conductive~\cite{Layani2009}, magnetic~\cite{Carstensen2018}, and other properties of these structures. Often, a transition from an amorphous zone to a dense packing (2D phase transition) is observed near the droplet periphery in these structures~\cite{Marin2011}, which is related to the competition between diffusive transport and the capillary flow induced by evaporation. In the central part of the deposit, some individual aggregates, such as chains and clusters of particles, can sometimes form, which is due to their capillary interaction~\cite{Park2006,Kolegov2019}. In the case of soft microgel particles, ordered monolayers of individually lying particles (loosely packed structures) can form~\cite{Jose2021}. Furthermore, many other factors can influence the structure: electrostatic interaction between particles~\cite{Molchanov2019}, magnetic properties of the materials~\cite{Carstensen2018}, Janus particles~\cite{Singh2025}, chemically structured substrates~\cite{Zhao2023}, and much more. To analyze such clusters, researchers use various characteristics: cluster size, particle packing density, fractal dimension, radius of gyration, etc.~\cite{Carstensen2018}. In the field of analyzing such structures, machine learning methods are now actively developing: a neural network can determine protein folding from an image of a dried biological fluid droplet~\cite{VelascoTeran2026}, predict the type and concentration of salt based on a crystalline deposit~\cite{Batista2025}, identify the presence of protein mutations (markers of certain diseases)~\cite{Jeihanipour2022}, detect biochemical changes for disease diagnosis based on the structures of dried blood droplets~\cite{Hamadeh2020}, and so on. For specialists working with colloidal systems, it would be extremely convenient to have the corresponding tools concentrated in one place. For this purpose, we previously created ISANM~\cite{Khusainova2025}, which is gradually being populated with various useful tools. A significant advantage is the ability to apply such tools simultaneously to a large set of images. However, before developing and using tools for analyzing colloidal images, it is highly desirable to first create modules capable of identifying individual particles and various types of their aggregates in the images. Subsequently, the obtained data can be exported to other modules for further analysis, which will help identify relationships between the morphology of colloidal structures and their physical properties.

Here we will focus on monolayers. Multilayer structures are not considered. We assume that all particles are of the same shape (spherical). We consider that in an image, particles can be of one size (with slight variation within the margin of error), or large and small particles may be present (binary mixtures~\cite{Inoue2020,Zolotarev2022}). Mixtures of particles of different shapes are not considered here.

\begin{figure*}[h]
	\center{\includegraphics[width=0.99\linewidth]{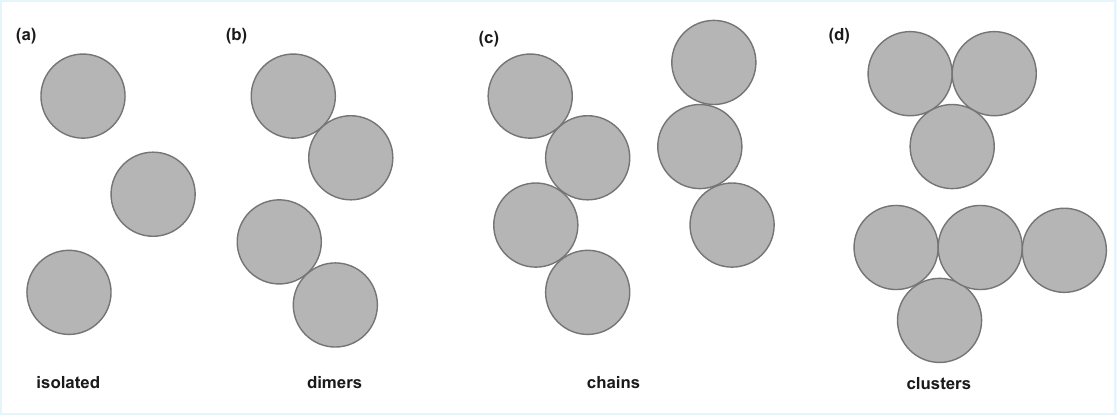}}
	\caption{Possible colloidal assemblies}
	\label{fig:Clusters}
\end{figure*}

Each assembly is characterized by a specific number of particles, which determines its category. In particular, the following assembly categories are distinguished~\cite{Lotito2020}:
\begin{itemize}[leftmargin=*, label=---]
	\item Isolated particles (particle $j$ belongs to this group if it has no neighbors within a distance $R_\mathrm{cut}$ from its center, see Fig.~\ref{fig:Clusters}a);
	\item Dimers (two particles $j$ and $k$ belong to this group if $j$ has only neighbor $k$ within a distance $R_\mathrm{cut}$ from its center and $k$ has only neighbor $j$ within a distance $R_\mathrm{cut}$ from its center, Fig.~\ref{fig:Clusters}b);
	\item Chains (in a chain, all particles have two neighbors, except for the particles at the two extremities of the chain, which have only one neighbor, Fig.~\ref{fig:Clusters}c);
	\item Clusters (particles form a cluster if they belong to an assembly that does not fall into any of the previous categories; this category also includes loops; particles belong to a loop if all of them have two neighbors within a distance $R_{\mathrm{cut}}$ from their centroid; a loop can be interpreted as a kind of ``closed'' chain or ring, Fig.~\ref{fig:Clusters}d).
\end{itemize}

Such types of assemblies can be of interest, for example, in tasks related to creating 1D and 2D photonic crystals~\cite{Shi2024}.

\subsection{Image analysis by threshold processing}
An important challenge addressed by colloidal analysis is the identification of particle assemblies (configurations)~\cite{Lotito2019}. In this study, image processing methods from the OpenCV library~\cite{DawsonHowe2014} were employed to analyze images of colloidal assemblies. The identification of assemblies is carried out in several stages:
\begin{itemize}[leftmargin=*, label=---]
	\item a threshold image (binarization) of particles is constructed, where particle neighborhoods are determined using particle centroids;
	\item to determine the neighborhood for each particle $j$, the set of neighbors is computed as the set $N_{j}$ of all particles located at a distance $R_{\mathrm{cut}}$ from the center of particle $j$ (the radius $R_{\mathrm{cut}}$ is calculated by the formula $R_{\mathrm{cut}} = 2 R + \epsilon$, where $\epsilon$ is a certain tolerance, calculated here by the formula $\epsilon = (r_1 + r_2)/10$, where $r_{1}$ and $r_{2}$ are the radii of two adjacent particles (since particle size may vary slightly in real images, $R = (r_1+r_2)/2$));
	\item the distance between two particles is calculated by the formula
	\[
	L = \sqrt{(x_{2} - x_{1})^2+(y_{2} - y_{1})^2},
	\] where $(x_{1},y_{1})$ are the coordinates of the centroid of particle $j$, $(x_{2},y_{2})$ are the coordinates of the centroid of another particle (if $L\leq R_\mathrm{cut}$, then the two particles are marked as neighbors);
	\item assemblies and their types (configurations) are determined based on the obtained information about the particles: their radii, centroid coordinates, and neighborhood relationships between particles.
\end{itemize}
It should be noted that the cut-off radius $R_\mathrm{cut}$ in our case is not a constant, as particle sizes can vary. It is continuously recalculated when checking the neighborhood of the current particle with others. A particle is considered a neighbor if its center lies within a distance $R_\mathrm{cut}$ from the center of the current particle. Since in the images we worked with, the number of particles did not exceed a thousand, the software implementation simply performs a full iteration over all particles, checking each particle for neighborhood with all others. However, in the future, if necessary (if the number of particles reaches several thousand or more), it is not difficult to add optimization to reduce computational costs. For example, by adding a spatial grid to search for a particle's neighbors only within its own cell and adjacent ones.

\begin{figure*}[htbp]
	\centering
	\includegraphics[width=0.5\linewidth]{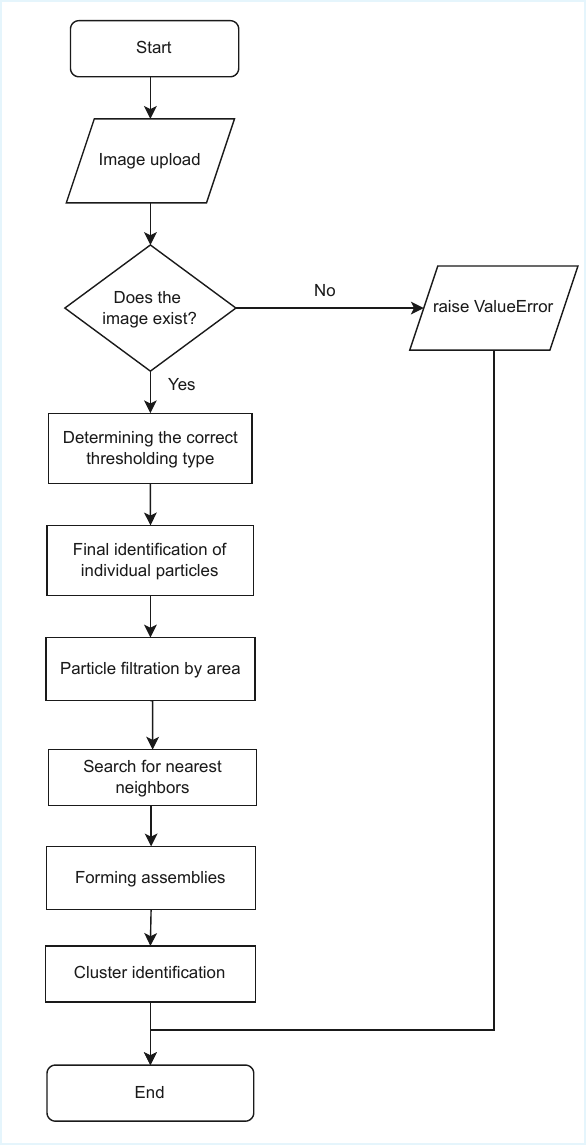}
	\caption{Algorithm flowchart}
	\label{fig:AlgorithmFrowchart}
\end{figure*}

\begin{figure*}[htbp]
	\centering
	\includegraphics[width=0.85\linewidth]{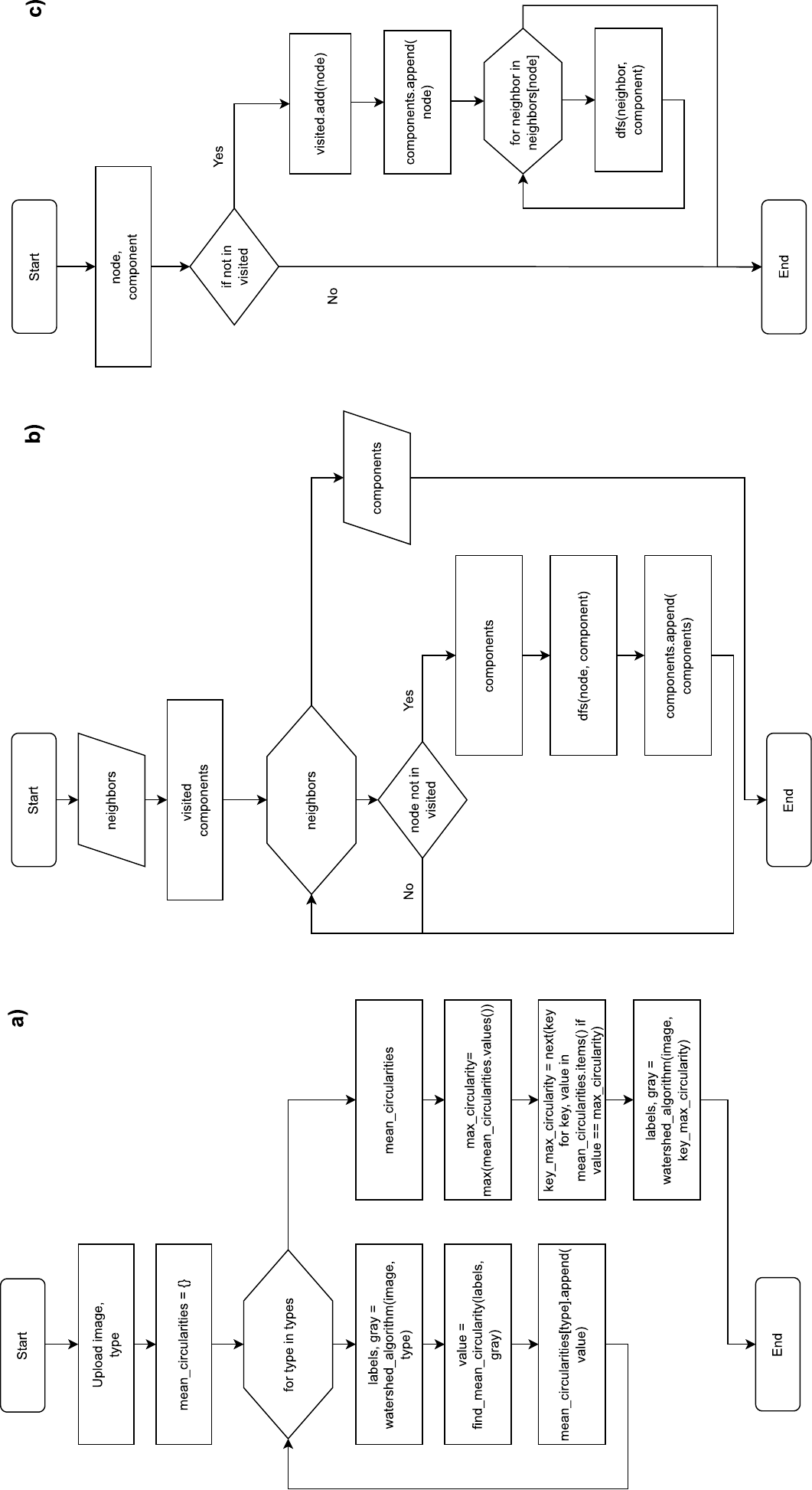}
	\caption{Subprocess flowcharts: a) flowchart of threshold type determination and particle identification, b) flowchart of assemblies formation, c) flowchart of dfs() function}
	\label{fig:SubprocessFlowcharts}
\end{figure*}

\begin{figure*}[htbp]
	\centering
	\includegraphics[width=0.9\linewidth]{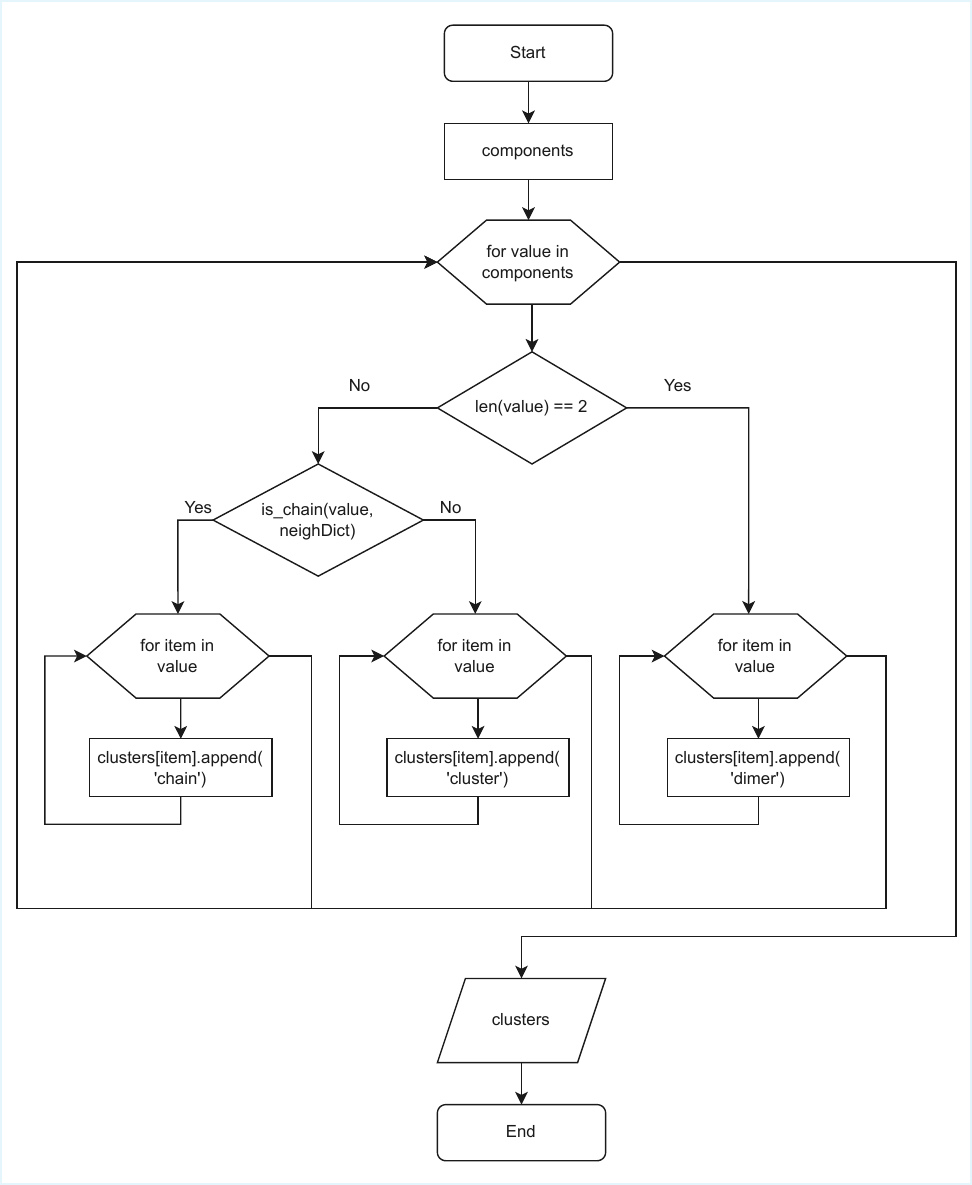}
	\caption{Flowchart of cluster identification process}
	\label{fig:Clustersidentification}
\end{figure*}

The overall flowchart of the image analysis process using the thresholding method is presented in Fig.~\ref{fig:AlgorithmFrowchart}. The algorithm includes the following key steps (detailed flowcharts of some key subprocesses are shown in Figs.~\ref{fig:SubprocessFlowcharts} and \ref{fig:Clustersidentification}):
\begin{enumerate}
	\item Image loading;
	\item Determining the correct thresholding type (Fig.~\ref{fig:SubprocessFlowcharts}a);
	\item Identification of individual particles (Fig.~\ref{fig:SubprocessFlowcharts}a);
	\item Particle filtering by area;
	\item Nearest neighbors search;
	\item Assembly formation (Figs.~\ref{fig:SubprocessFlowcharts}b and \ref{fig:SubprocessFlowcharts}c);
	\item Assembly identification (Fig.~\ref{fig:Clustersidentification}).
\end{enumerate}

The array storing the mean circularity values of the particles is denoted as \texttt{mean\_circularities} (Fig.~\ref{fig:SubprocessFlowcharts}a). The maximum circularity value from this array is stored in the variable \texttt{max\_circularity}. The type of thresholding is specified in the variable \texttt{key\_max\_circularity}. In Figures~\ref{fig:SubprocessFlowcharts}b and \ref{fig:SubprocessFlowcharts}c, ``\texttt{components}'' is a list of all found connected components (particle assemblies). The \texttt{dfs()} function implements the Depth First Search (DFS) method. Connections between particles are described using a neighbor list \texttt{neighDict} (Fig.~\ref{fig:Clustersidentification}). Based on the number of neighbors of particles, an analysis of connected components is performed, allowing the distinction between chains and clusters.

Image binarization is the process of converting a color image into a monochrome (black and white) image using a specified threshold value. Pixels whose intensity exceeds the set threshold are assigned white (encoded as 1), while pixels with values not exceeding the threshold are set to black (encoded as 0). This is a key step in many computer vision tasks. However, the main difficulty lies in selecting the optimal threshold that correctly separates pixels into objects and background, especially in the presence of noise, lighting gradients, and complex structures. Due to the ambiguity and heterogeneity of real images of colloidal assemblies, automatic threshold selection is critically important, capable of effectively separating pixels across all image types. Such an algorithm must adapt to variations in brightness, contrast, and artifacts, ensuring correct data processing throughout the entire image dataset. The OpenCV library offers two approaches to automate the process: global thresholding (Otsu's method) and adaptive (local) thresholding.

The essence of adaptive thresholding is that the algorithm determines the threshold for a pixel based on a small region surrounding it. Thus, different thresholds are obtained for different areas of the same image, which yields better results for images with varying illumination.

The following methods are used to calculate the threshold value:
\begin{enumerate}
	\item cv2.ADAPTIVE\_THRESH\_MEAN\_C (the threshold value is the mean of the neighborhood area minus a constant $C$);
	\item cv2.ADAPTIVE\_THRESH\_GAUSSIAN\_C (the threshold value is the Gaussian-weighted sum of the neighborhood area values minus a constant $C$).
\end{enumerate}

This approach is not optimal for the present work because, despite the automatic calculation of the threshold value, there are parameters that need to be set manually: \textbf{blockSize} defines the size of the moving window's neighborhood, and $C$ is a constant subtracted from the calculated threshold value. Depending on the image, these parameters may need to be adjusted.

Otsu's method is an algorithm for automatically determining the optimal binarization threshold, based on a statistical analysis of the image histogram. It calculates a threshold value that maximizes the inter-class variance (the difference between the two pixel groups) and minimizes the intra-class variance (the spread of values within each group). Thus, the method seeks a threshold at which the difference between the mean values of the two classes (dark and light pixels) is maximized, while simultaneously reducing the variation among pixels within each class, thereby increasing the homogeneity of the segmented regions~\cite{Otsu1979}. This method is the most optimal for the given task, as it does not require manual parameter adjustment.

Ultimately, Otsu's algorithm was chosen as the image binarization method, demonstrating the highest effectiveness for the given task. However, our tests  showed that thresholding provides only a generalized visualization of the assembly structure. Solving the classification problem requires a more detailed separation of all particles (there should be a small gap between them), which necessitates a modification of the approach. One of the image segmentation methods can assist in this. It is the Watershed algorithm~\cite{Vincent1991}.

To understand how the Watershed algorithm works, one should imagine a grayscale image as a topographic surface, where high-intensity pixel values (lighter areas) represent peaks, and low-intensity pixels (darker areas) represent depressions (local minima). If we start filling the depressions with water, the water will soon begin to overflow from the edges. For this purpose, so-called watershed lines are constructed. These lines separate objects in the image, preventing them from merging.

Furthermore, during the work, a problem with the non-uniformity of the input data was identified: in some cases, particles are brighter than the background (e.g., light objects on a dark background), while in others they are darker than the background (dark objects on a light background). Applying the thresholding types \texttt{THRESH\_BINARY} and \texttt{THRESH\_BINARY\_INV} for individual manual threshold adjustment for each image in large datasets is inefficient, as it requires significant time investment.

THRESH\_BINARY: if the intensity of a pixel exceeds the set threshold, it is assigned a value of 255 (white), otherwise it is assigned a value of 0 (black)~\cite{Gonzalez2018},
\[
dst(x, y) =
\begin{cases}
	maxval, & \text{if } src(x, y) > thresh, \\
	0,            & \text{otherwise},
\end{cases}
\]
where $dst(x, y)$ is the value of the pixel at coordinates $(x,y)$ in the transformed image, $src(x,y)$ is the value of the pixel at coordinates $(x,y)$ in the source image, $thresh$ is the threshold value (a value from 0 to 255 determined automatically using Otsu's method), $maxval$ is the maximum value assigned to pixels exceeding the threshold (usually 255 for white).

It should be noted that in classical literature on image processing, for example in~\cite{Gonzalez2018}, a normalized representation of pixel intensities from 0 to 1 is often used for theoretical analysis, where $maxval = 1$. In the present work, for convenience and practical implementation, an integer representation with a maximum value of 255 is used.

THRESH\_BINARY\_INV: the algorithm is opposite to THRESH\_BI\-NARY,
\[
dst(x, y) =
\begin{cases}
	0, & \text{if } src(x, y) > thresh,\\
	maxval,            & \text{otherwise}.
\end{cases}.
\]

To automate the selection of the threshold type depending on the input image, the circularity parameter~\cite{Gonzalez2018} was used. It is calculated as
\[
\text{circularity} = \frac{4\pi A}{P^2},
\]
where $P$ is the perimeter of the particle and $A$ is the area of the particle. For ideal circles, the circularity parameter tends to 1. When analyzing identified particles, the circularity values obtained with different thresholding types are compared. The method that provides the maximum average circularity value is selected as the optimal one (Fig.~\ref{fig:SubprocessFlowcharts}a).

After the successful identification of individual particles belonging to a particular configuration (assembly), the nearest neighbors are determined for each of them. However, the key task remains the correct identification of all assemblies formed based on local connections between particles, as they reflect the global structure in the image. To solve this problem, the Depth First Search (DFS) algorithm~\cite{Cormen2009} was applied. The algorithm operated according to the following principle (Fig.~\ref{fig:SubprocessFlowcharts}c):
\begin{enumerate}
	\item Select any particle that has not yet been visited.
	\item Execute the \texttt{dfs()} procedure, in which:
	\begin{enumerate}[label*=\arabic*.]
		\item Mark the particle as visited.
		\item For each unvisited neighboring particle, execute the \texttt{dfs()} procedure.
	\end{enumerate}
	\item Repeat steps 1 and 2 until all particles have been visited.
\end{enumerate}

\subsection{Machine learning}
\subsubsection{Dataset}
Datasets play a significant role in machine learning. However, existing datasets are not always suitable for highly specialized tasks. In such cases, one has to create their own dataset for specific tasks.

\begin{figure}[h]
	\centering
	\includegraphics[width=0.9\linewidth]{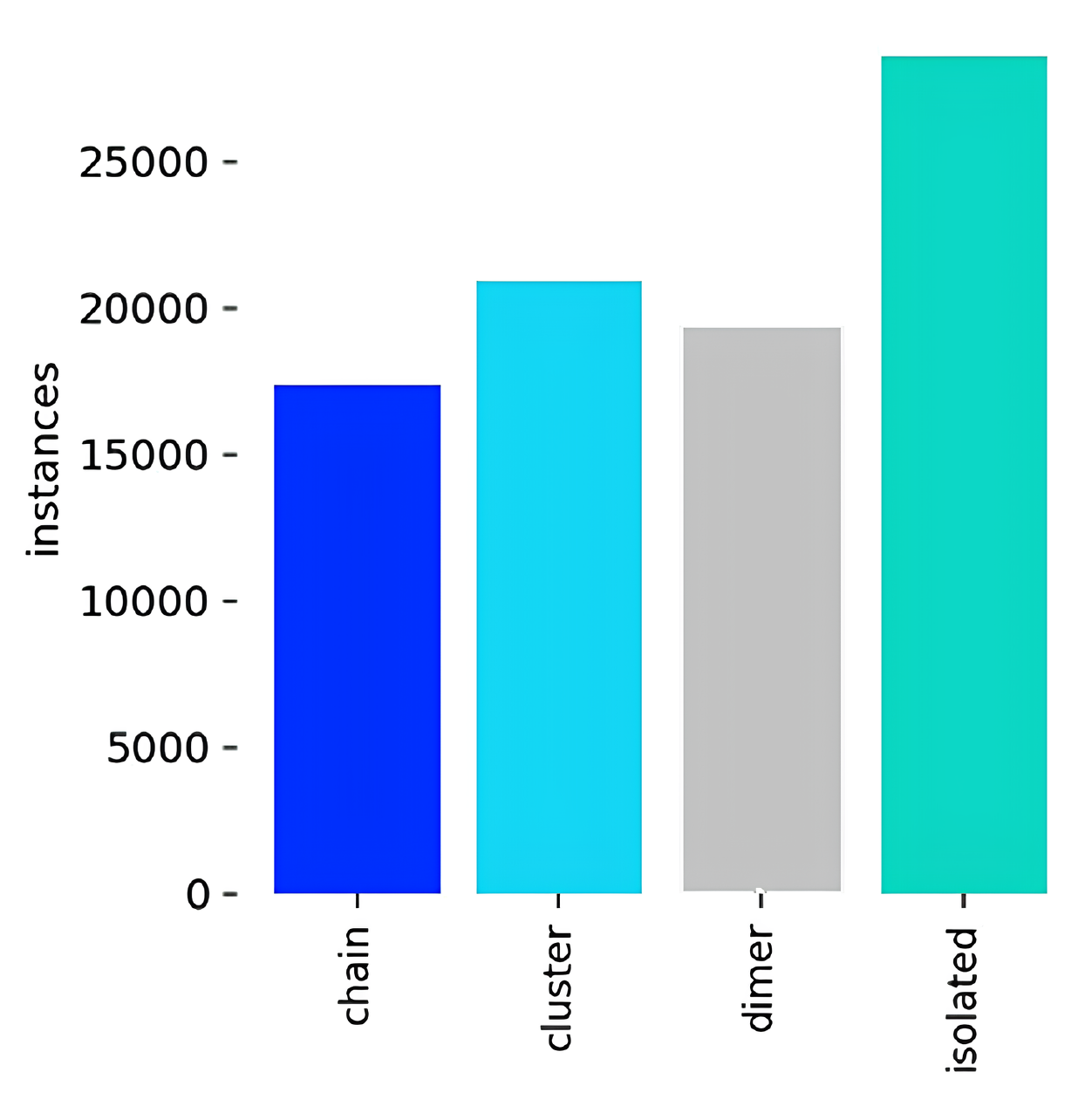}
	\caption{Number of class labels}
	\label{fig:Labels}
\end{figure}

In the course of work on the given task, a dataset has been created, currently consisting of over six thousand images (experimental and simulation-based). The proportion of simulated images in this dataset is small (approximately 6.5\%). Thus, the dataset predominantly consists of experimental images taken from numerous scientific papers. Such images are typically obtained using SEM (Scanning Electron Microscope) and TEM (Transmission Electron Microscope) methods. Simulated images were generated using algorithms described in previous works~\cite{Zolotarev2022,Zolotarev2021}. The dataset was annotated manually using the Roboflow service~\cite{RoboFlow2025}. This dataset was automatically and randomly split into three subsets in the following ratio: training~--- 80\%, validation~--- 10\%, and test one~--- 10\%. Each data subset consists of original images (no data leakage occurred) and their corresponding labels (Fig.~\ref{fig:Labels}). An example of such labels is presented in Table~\ref{tab:DataStructure}.

\begin{table*}
	\caption{Structure of data records}
	\centering
	\begin{tabular}{|c|c|c|c|c|c|}
		\hline
		Class & $x_1$ & $y_1$ & $x_2$ & $y_2$ & \dots \\ \hline
		2 & 0.0971293 & 0.599966 & 0.0971293 & 0.6136681 & \dots \\
		1 & 0.7138894 & 0.7144959 & 0.8073184 & 0.7137358 & \dots \\
		3 & 0.9428706 & 0.6332427 & 0.9487924 & 0.6381358 & \dots \\
		1 & 0.4452560 & 0.6059420 & 0.4393396 & 0.6107149 & \dots \\
		2 & 0.7403180 & 0.5285189 & 0.7379487 & 0.5285189 & \dots \\ \hline
	\end{tabular}
	\label{tab:DataStructure}
\end{table*}

The structure of the data records is as follows: first comes the class number to which the created polygon belongs. The created dataset contains 4 labels: 0~--- chain, 1~--- cluster, 2~--- dimer, 3~--- isolated particle. After the labels, the normalized coordinates of the polygon points are listed. Their number can vary, depending on how many points were placed on the polygon during labeling.

The dataset was compiled from images of various sizes. To ensure uniformity, all images were resized to a standard size of $640\times 640$ pixels using the Roboflow toolkit. Image augmentation was applied to increase data diversity and improve the model's generalization capability. The dataset size was doubled (from 3324 to 6648 images) by applying the following filters: Deblur, Gaussian Blur, Salt and Pepper Noise, Brightness Adjustment, Contrast Adjustment. This increased the model's accuracy from approximately 70\% to 97\%.

\begin{figure}
	\centering
	\includegraphics[width=0.99\linewidth]{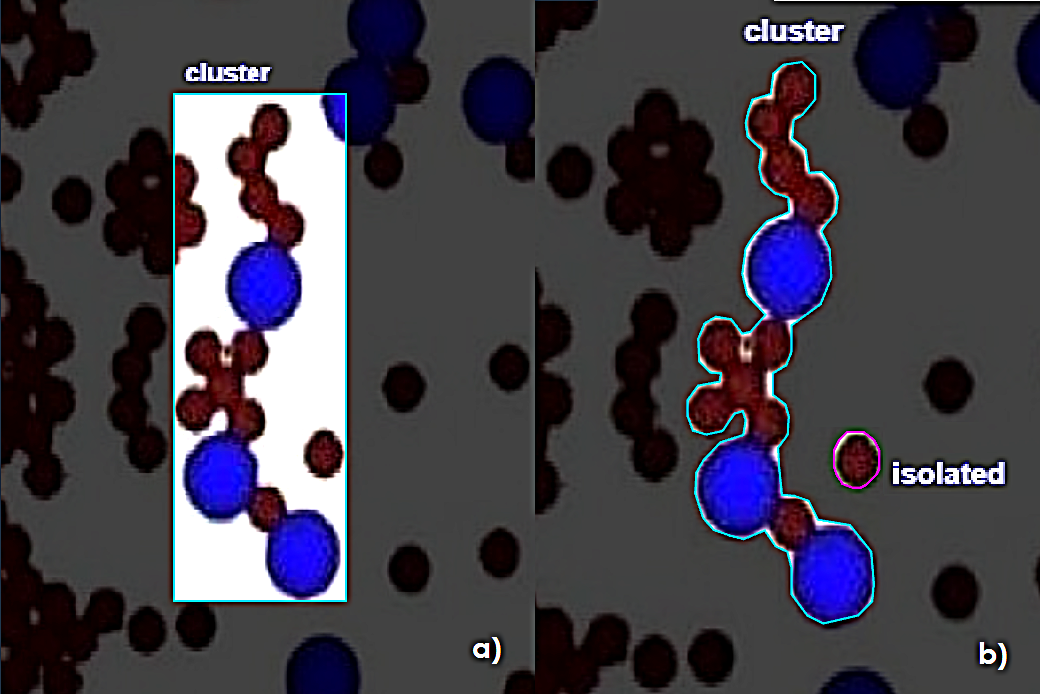}
	\caption{Two types of data annotation: (a) bounding boxes and (b) polygons (the image was generated using an algorithm from a previous study~\cite{Zolotarev2022})}
	\label{fig:BoundingBoxesVsPolygons2}
\end{figure}

As mentioned earlier, the structure of the data records consists of polygon points instead of the conventional coordinates of bounding box corners. The reason is that the method of object localization using polygons (Instance segmentation technology) was employed instead of the method using ``bounding boxes'' (Object detection technology). This was necessitated by some images where different assemblies were located at such minimal distances from each other that the use of bounding boxes was impractical, as a part of one assembly would fall into the area of a bounding box highlighting a completely different assembly (Fig.~\ref{fig:BoundingBoxesVsPolygons2}). Thus, it was concluded that using the object localization method with ``bounding boxes'' might be insufficient for delineating the structure of complex objects. In turn, object localization using polygons is suitable for working with complex-shaped objects, especially overlapping or intertwined ones, as colloidal assemblies can exhibit interesting and unpredictable structures.

\subsubsection{Model YOLOv8}
\begin{figure*}[h]
	\centering
	\includegraphics[width=0.98\linewidth]{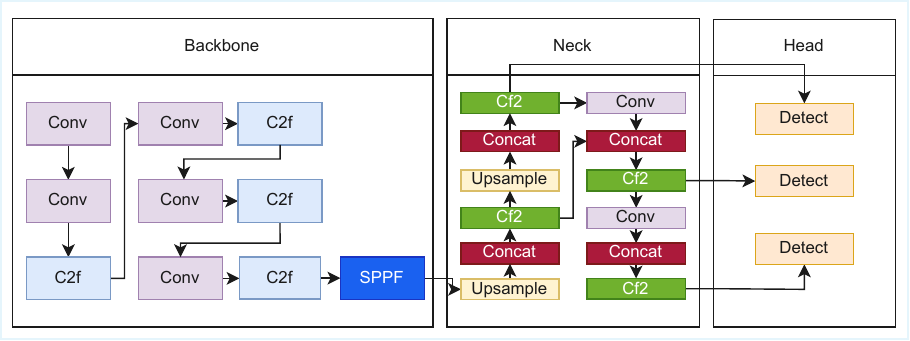}
	\caption{Model Architecture YOLOv8 (Yaseen, Licence CC BY 4.0)\cite{yaseen2024yolov8indepthexplorationinternal}}
	\label{fig:YOLOv8}
\end{figure*}

Furthermore, a machine learning method was also used as one of the approaches for identifying colloidal assemblies in images. Our choice settled on the YOLOv8 model. This model (the model architecture is presented in Fig.~\ref{fig:YOLOv8}) possesses not only high speed but also high accuracy in object detection and recognition, and it supports the full spectrum of computer vision tasks, including the segmentation used in this study.

The YOLOv8 model consists of three components: `backbone', `neck', and `head'. The `backbone' serves to extract features from the input image and is typically a deep convolutional neural network. The `neck' processes the features extracted by the `backbone' and may include additional layers and mechanisms to refine the extracted features. The `head' makes the final predictions based on the previous parts of the architecture. The SPPF (Spatial Pyramid Pooling-Fast) layer integrates information about object features at different scales (Fig.~\ref{fig:YOLOv8}). It applies several pooling operations with different kernel sizes, enabling the network to simultaneously ``see'' both small and large objects in the image. The Conv (Convolutional Layer) applies filters to the image to extract edges, textures, and other features. C2f (Cross Stage Partial network with 2 convolutions + fusion) is an enhanced feature extraction block. It splits the input features into two parts. One part is then passed through convolutions and subsequently merged with the other part to save computational resources while preserving feature quality. Concat (Concatenation) is a layer that merges multiple inputs together, allowing the network to see all the information simultaneously. Upsample is a layer that increases the size of the feature map. It is used to enable the network to work with smaller objects in the image. 

It is worth noting that the YOLOv8 model is pre-trained on large datasets, for example, COCO (Common Objects in Context)~\cite{lin2014microsoft} for computer vision tasks. COCO contains over 330~000 images with 80 object classes (e.g., person, car, etc.). Such pre-training allows the model to ``know'' basic object features, such as shape, texture, etc. However, for highly specialized tasks, this is insufficient, and fine-tuning on a custom dataset is required. YOLOv8 has different size variants: YOLOv8n, YOLOv8s, YOLOv8m, YOLOv8l, and YOLOv8x. Smaller models provide high speed, while larger ones, in turn, provide deeper processing and better results.

For segmentation tasks, the YOLOv8 model was adapted into YOLOv8-Seg with corresponding size variants: YOLOv8n-seg, YOLOv8s-seg, YOLOv8m-seg, YOLOv8l-seg, and YOLOv8x-seg. The architecture of the YOLOv8-Seg model is similar to the YOLOv8 architecture, but adds two key components: a Proto Module (FCN layer~--- Fully Convolutional Network) for generating mask prototypes and a module for predicting mask coefficients. These coefficients are combined with the prototypes to produce final pixel-wise masks for each object. The hyperparameters presented in Table~\ref{tab:ModelHyperparameters} were used for training and validating the model.

\begin{table}[h]
	\centering
	\caption{Hyperparameters of the model}
	\begin{tabular}{|l|l|}
		\hline
		\textbf{Parameter} & \textbf{Value} \\ \hline
		Epochs & 100 \\
		Image size & 640 \\
		Batch size & 35 \\
		Optimizer & AdamW \\
		Learning rate & 0.00125 \\
		Momentum & 0.937 \\
		Label smoothing & 0.2 \\
		Activation function & SiLU \\
		Loss & BCE Loss, DFL+CIou Loss \\ \hline
	\end{tabular}
	\label{tab:ModelHyperparameters}
\end{table}

It is worth noting that the values of parameters such as `batch size', `optimizer', and `momentum' were selected automatically by the YOLOv8 model as the most optimal during the training phase. The parameter `label\_smoothing' was also manually selected and is used to regulate the model's training process. This parameter helps prevent the model from becoming overconfident in its predictions by slightly smoothing the target values, thereby reducing the likelihood of overfitting.

The following metrics are used as measures for evaluating the quality of the model training: `mAP', `Precision', and `Recall'. `Precision' shows how accurate the model is in its predictions. In other words, this value represents the proportion of correctly detected assemblies among all assemblies that the model considered correct,
\[
\text{Precision} = \frac{\mathrm{TP}}{\mathrm{TP} + \mathrm{FP}},
\] where $\mathrm{TP}$ is the number of true positive objects, i.e., objects correctly assigned to the class, and $\mathrm{FP}$ is the number of false positive objects, in other words, objects incorrectly assigned to the class.

`Recall' shows how completely the model identifies all assemblies present in the image. In other words, this value represents the proportion of correctly detected assemblies among all objects that should be found,
\[
\text{Recall} = \frac{\mathrm{TP}}{\mathrm{TP} + \mathrm{FN}},
\] where $\mathrm{FN}$ is the number of false negative objects, for which it is incorrectly stated that the objects do not belong to the class.

The `mAP' metric is the main metric for evaluating object identification models. This metric combines the aforementioned `Precision' and `Recall', and also shows how well the model performs on average for all object classes,
\[
\text{mAP} = \frac{1}{N} \sum_{i=1}^{N} \mathrm{AP}_i,
\] where $N$ is the number of classes, and $\mathrm{AP}_i$ is the Average Precision for class $i$ (calculated as the area under the Precision-Recall curve).

\section{Results and discussions}

During the work on particle assembly recognition using image analysis through threshold processing, it was decided that adding interactive sliders would facilitate the construction of the threshold image, allowing the user to observe in real-time how the threshold image changes depending on the selected parameters of the HSV color model. This interactivity enables the user to independently select optimal settings, contributing to more accurate and adaptive object segmentation in the image. The developed tools have been implemented into the Information System for Analysis of Nanostructure Morphology (ISANM)~\cite{Khusainova2025}.

When dealing with the processing of a large number of images, it is necessary to automate the process of selecting optimal parameters, as manual adjustment in this case would be too time-consuming. In the developed program~\cite{KhusainovaThresholding2025} for image analysis using the thresholding method, to apply the Watershed algorithm, a threshold image is first constructed using Otsu's method. The Watershed algorithm requires a monochrome image, so the original image is converted to grayscale. Additionally, to minimize false contours, image smoothing is performed prior to segmentation using Gaussian blur.

\begin{figure}[htbp]
	\centering
	\includegraphics[width=0.8\linewidth]{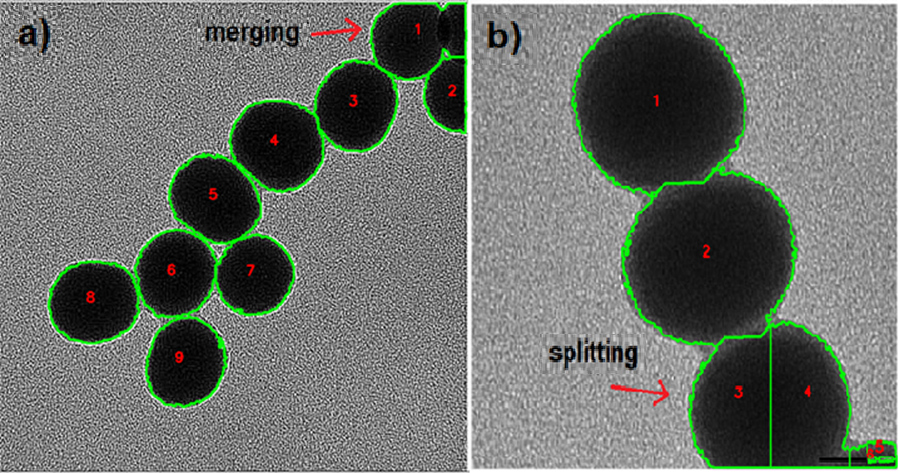}
	\caption{Incorrect interpretation of borders (TEM images): a) merging, b) splitting (Haddadi et al., Licence CC BY 4.0) \cite{Haddadi2021}.}
	\label{fig:MerginSplitting}
\end{figure}

Despite the overall functionality of the method, segmenting particle contours using the Watershed algorithm can be accompanied by local artifacts in the form of irregularities (roughness may occur at particle edges due to the algorithm's sensitivity to brightness variations) and partial merging (in areas of dense particle arrangement, boundaries of adjacent objects are sometimes interpreted incorrectly, leading to their merging (Fig.~\ref{fig:MerginSplitting}a) or splitting of a particle (Fig.~\ref{fig:MerginSplitting}b)).

\begin{figure}[htbp]
	\centering
	\includegraphics[width=0.49\linewidth]{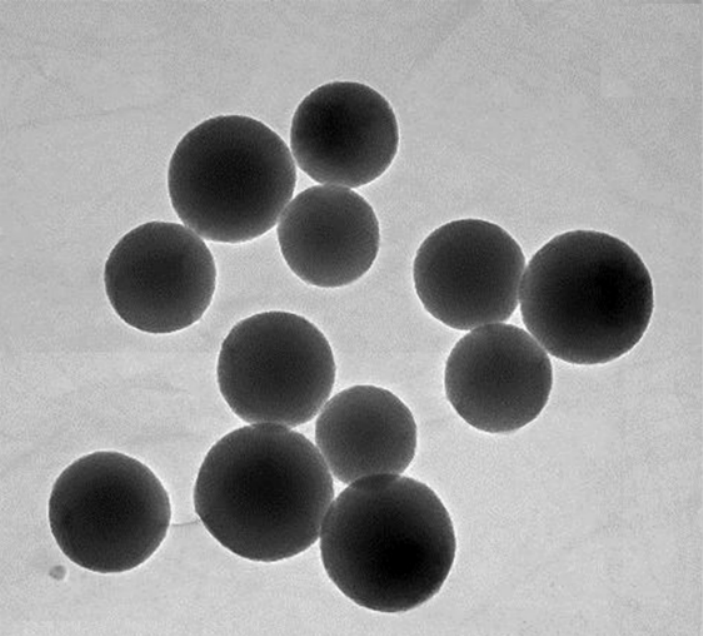}(a)
	\includegraphics[width=0.49\linewidth]{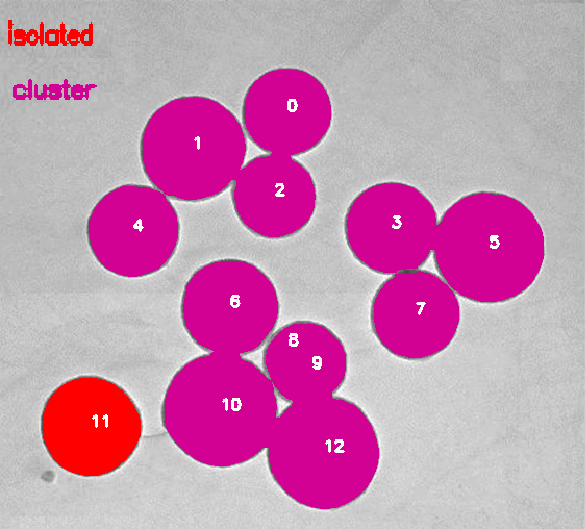}(b)
	\caption{(a) The original TEM image (Copyright Elsevier, 2016)~\cite{Kurdyukov2016} and (b) assembly recognition result.}
	\label{fig:IdentificationResult}
\end{figure}

Each category of identified assembly is marked with its own color for convenience: isolated particles are marked in red, chains in green, dimers in yellow, and clusters in pink. The final result of particle assembly recognition using image analysis by the thresholding method is presented in Fig.~\ref{fig:IdentificationResult}b (the original image is shown in Fig.~\ref{fig:IdentificationResult}a).

\begin{figure}[htbp]
	\centering
	\includegraphics[width=0.99\linewidth]{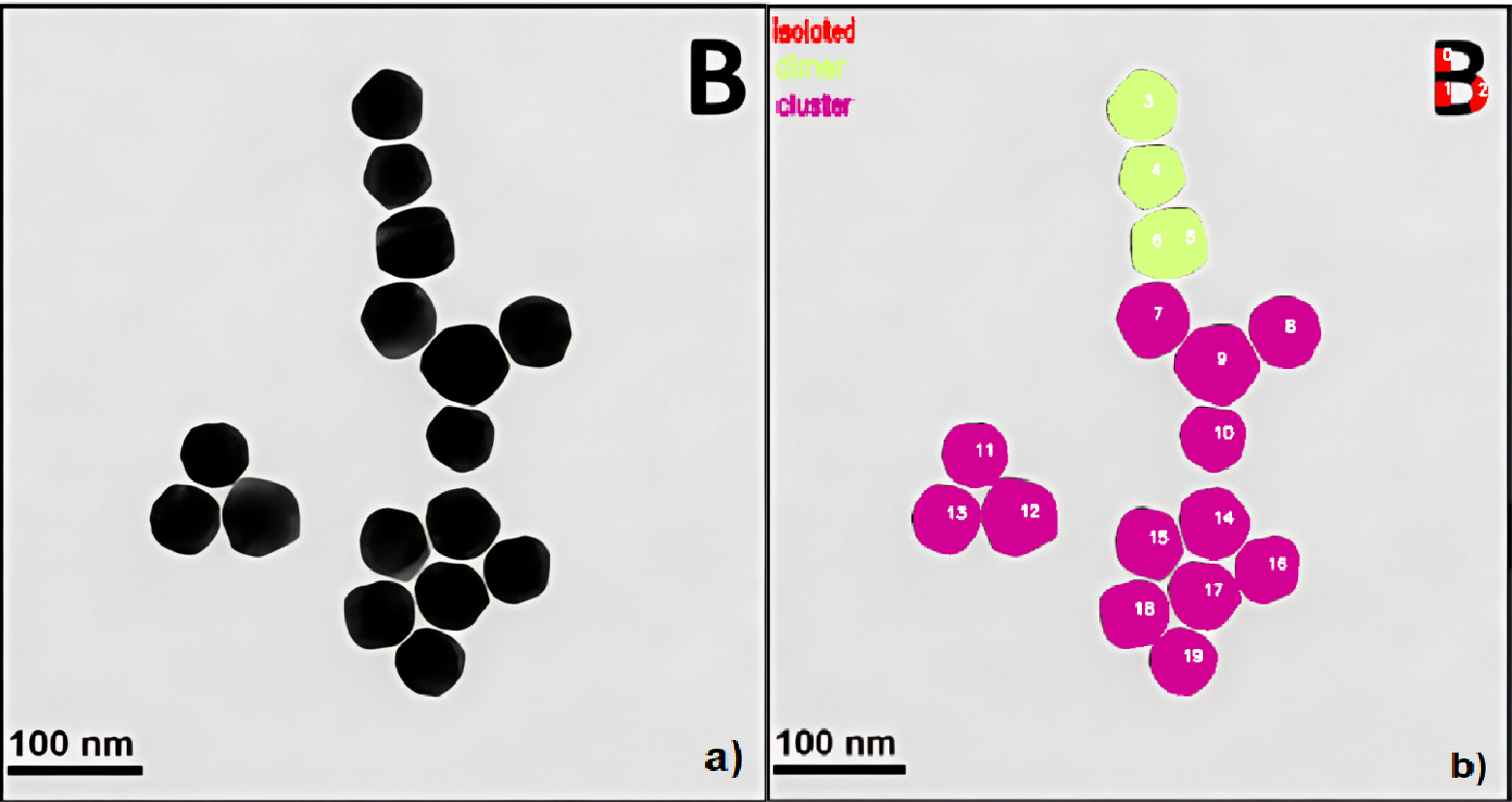}
	\caption{Recognition result of the unprocessed image: a) original TEM image (Shilo et al., Licence CC BY 4.0)~\cite{Shilo2015}, b) assembly recognition result.}
	\label{fig:RawImageResult}
\end{figure}

It is important to emphasize that the thresholding method here was applied exclusively to images that underwent a preprocessing stage: text labels, noise, and artifacts were manually removed from the images. Neglecting this preparation would lead to the false identification of such interferences as particles, which would significantly reduce the analysis accuracy. But this does not provide an absolute guarantee; for example, in Fig.~\ref{fig:IdentificationResult}b, one of the particles is split into two (8 and 9). Figure~\ref{fig:RawImageResult} demonstrates the result of thresholding applied to an image without preliminary filtering and manual processing (the letter ``B'' was intentionally preserved in the source image). It can be seen that the text label was erroneously identified as three single particles. This highlights the necessity of the preprocessing stage. Furthermore, part of a cluster was falsely identified as two dimers, and one particle was falsely recognized as two adjacent ones. This is likely related to the shape of the particles and the specific characteristics of the identification algorithm.

Machine learning, in turn, enables the automatic extraction of features from data, which is particularly useful for complex images with heterogeneous structures. This method can adapt to various conditions (such as changing lighting or particle shapes), making it more flexible compared to the thresholding (binarization) method.

\begin{figure}[htbp]
	\centering
	\includegraphics[width=0.99\linewidth]{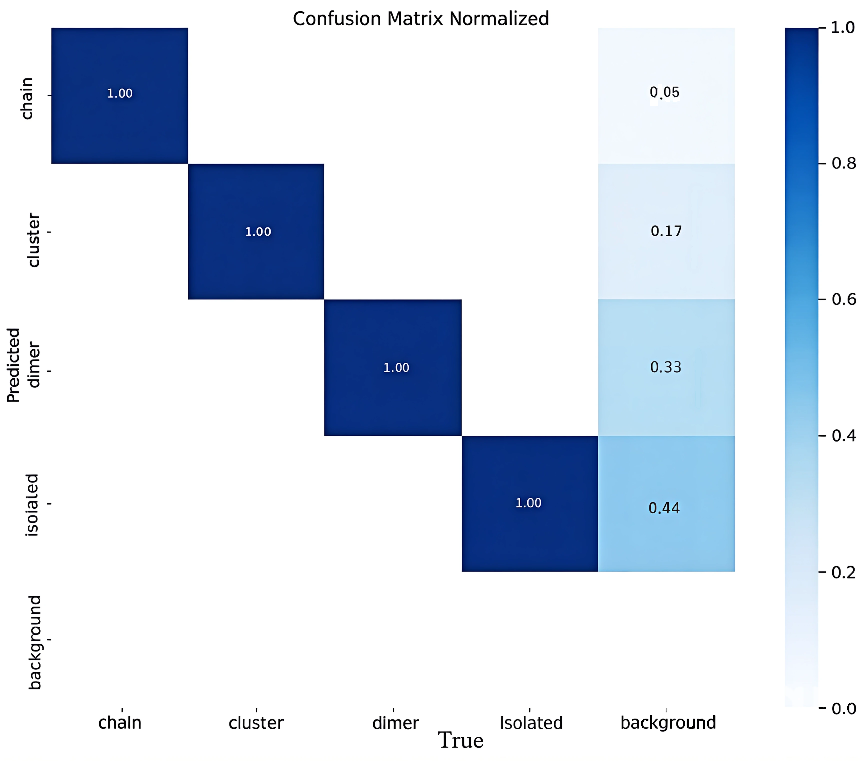}
	\caption{Error matrix.}
	\label{fig:ConfusionMatrix}
\end{figure}

The YOLOv8-Seg model was trained on a server with an AMD EPYC 7713 64-core Processor and 8 NVIDIA Quadro RTX A6000 GPUs. Within the scope of this work, considering the size of the created dataset, we trained various YOLOv8-Seg architectures: from the compact YOLOv8n-Seg to the balanced YOLOv8m-Seg. Based on the evaluation of accuracy metrics (mAP) and training speed, the YOLOv8m-Seg model demonstrated the best balance between efficiency and assembly identification quality. As a result of the model training, a confusion matrix was created (Fig.~\ref{fig:ConfusionMatrix}), showing the frequency of classification errors. As can be seen from the figure, the model shows almost perfect accuracy for all classes. However, there is a slight probability of misclassifying objects as background. The highest error probability is observed for isolated particles. Presumably, this is because the model may misinterpret noise as isolated particles.

\begin{figure*}[htbp]
	\centering
	\includegraphics[width=0.99\linewidth]{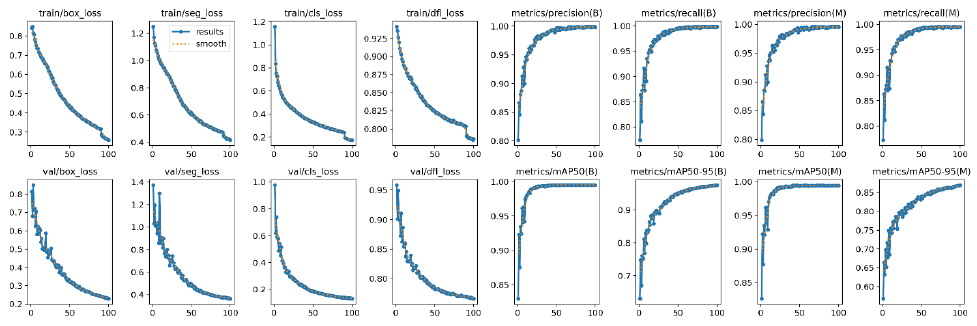}
	\caption{The quality of the model's training.}
	\label{fig:LearningCurves}
\end{figure*}

Furthermore, to assess the quality of the model training process, the following part of this section presents plots showing the changes in various evaluation metrics and losses during the model's training (Fig.~\ref{fig:LearningCurves}). These plots help evaluate how well the model is learning and identify potential issues.

The model demonstrates high-quality object detection and segmentation, as evidenced by the mAP50, Precision, and Recall values. The successful training of the model is indicated by the losses, which consistently decreased with each epoch. The best results were achieved at the 100th epoch, making this point optimal for using the model. Since the model had already achieved high results by the 100th epoch, further training may be unjustifiably costly.

\begin{figure}[htbp]
	\centering
	\includegraphics[width=0.99\linewidth]{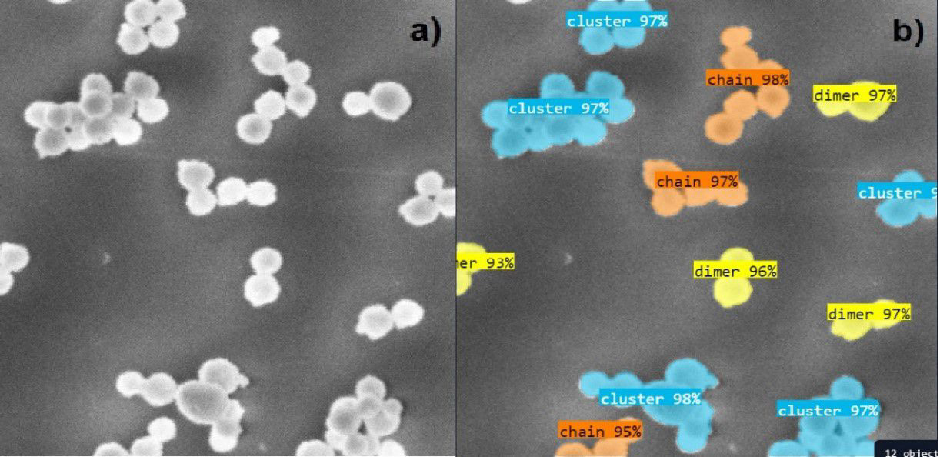}
	\caption{Prediction result of the model: a) original SEM image (Alqedra et al., Licence CC BY 4.0)~\cite{Alqedra2023}, b) analysis result.}
	\label{fig:ModelResult}
\end{figure}

Figure~\ref{fig:ModelResult} presents the model's prediction results. The model demonstrates high effectiveness when working with images from the test dataset, which were not used during the training process. This indicates the model's good generalization ability, meaning it successfully applies the acquired knowledge to new, previously unseen data.

\begin{figure}[htbp]
	\centering
	\includegraphics[width=0.99\linewidth]{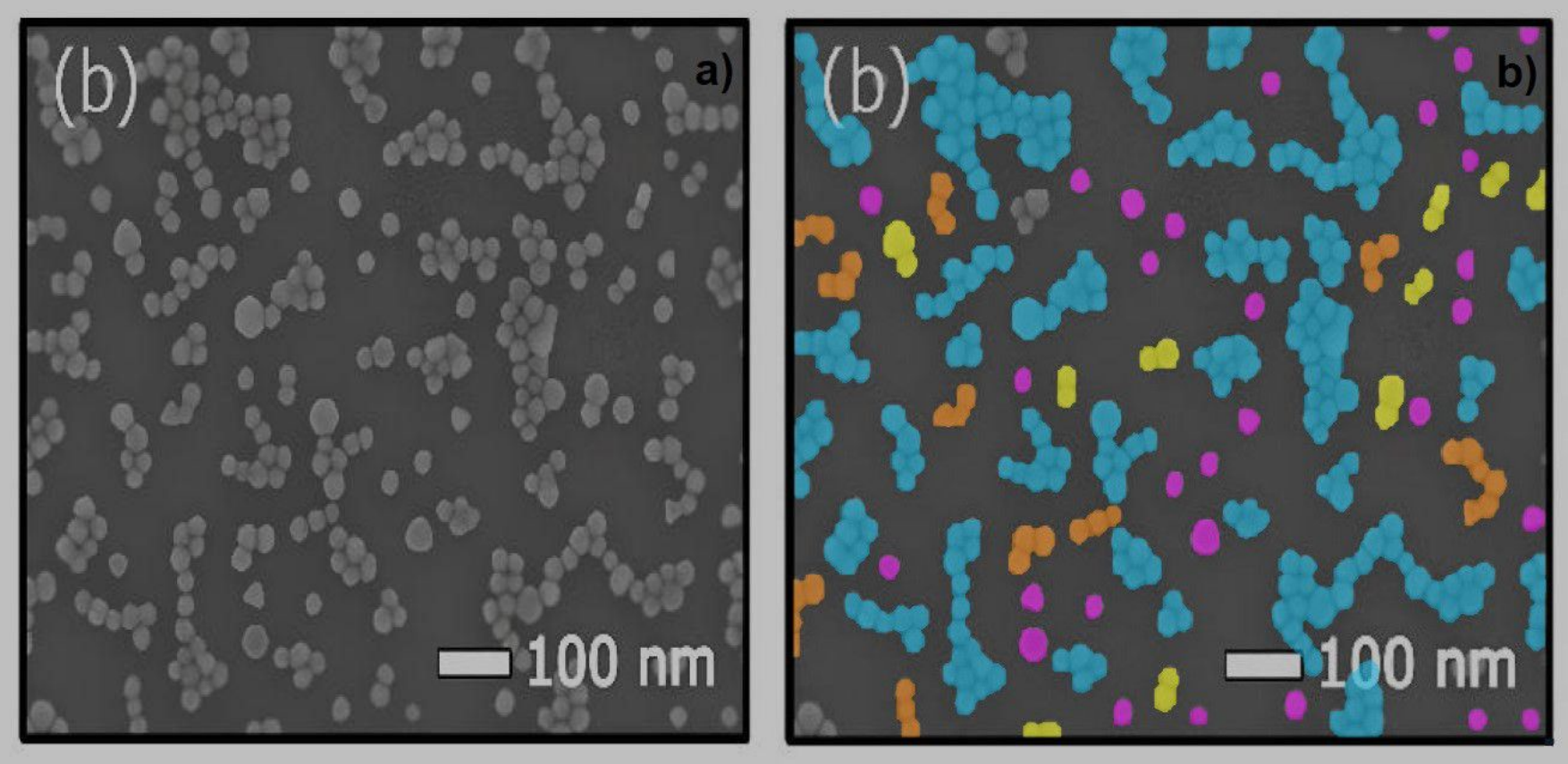}
	\caption{Prediction result of the model: (a) original SEM image (Poghossian et al., Licence CC BY 4.0)~\cite{Poghossian2022}, (b) analysis result (isolated particles are marked in pink, chains in brown, dimers in yellow, and clusters in blue).}
	\label{fig:High?oncentrationResult}
\end{figure}

The analysis of the results presented in Fig.~\ref{fig:High?oncentrationResult} demonstrates a decrease in the effectiveness of particle recognition in areas of their high concentration: individual objects remain unidentified. This phenomenon is likely due to the data annotation method: particles located in the edge regions of the image were not annotated due to the ambiguity of their cluster configuration, which was truncated by the image border. This annotation approach possibly led to the systematic omission of some edge assemblies and not only those. All of the above forms a basis for further research. It is worth noting that, unlike the thresholding method, inscriptions (labels) on the figures do not create problems for the machine learning method when identifying assemblies with spherical particles, as they are simply ignored.

\begin{figure*}
	\begin{minipage}[h]{0.3\linewidth}
		\center{\includegraphics[width=0.8\linewidth]{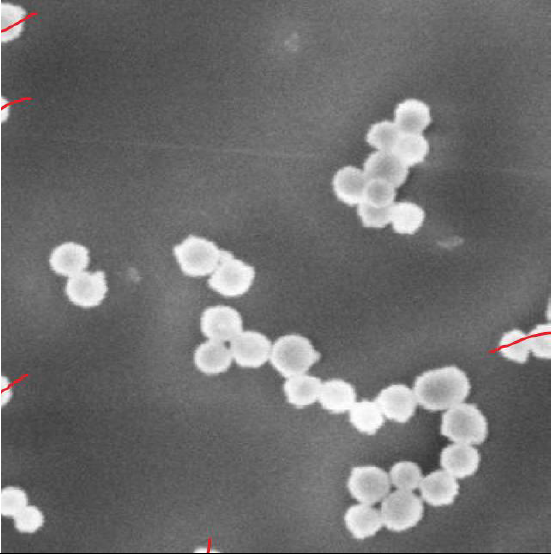}(a)}
	\end{minipage}
	\begin{minipage}[h]{0.3\linewidth}
		\center{\includegraphics[width=0.8\linewidth]{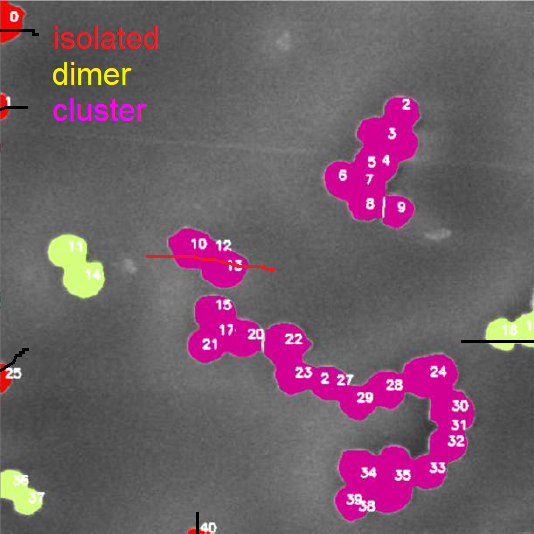}(b)}
	\end{minipage}
	\begin{minipage}[h]{0.3\linewidth}
		\center{\includegraphics[width=0.8\linewidth]{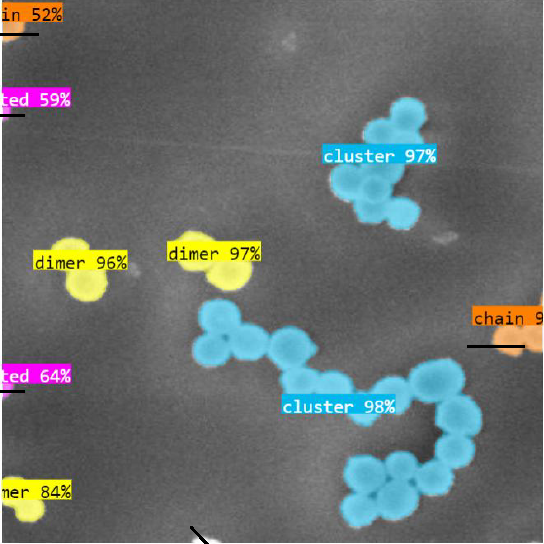}(c)}
	\end{minipage}
	\\
	\begin{minipage}[h]{0.3\linewidth}
		\center{\includegraphics[width=0.8\linewidth]{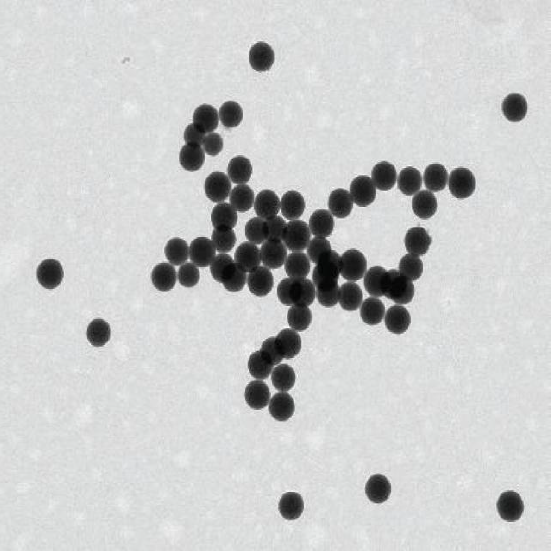}(d)}
	\end{minipage}
	\begin{minipage}[h]{0.3\linewidth}
		\center{\includegraphics[width=0.8\linewidth]{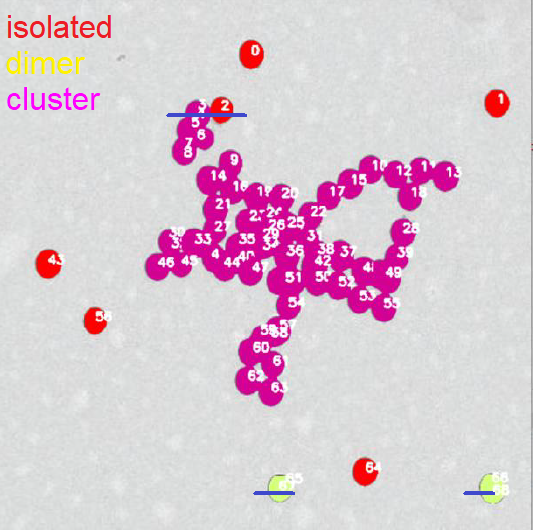}(e)}
	\end{minipage}
	\begin{minipage}[h]{0.3\linewidth}
		\center{\includegraphics[width=0.8\linewidth]{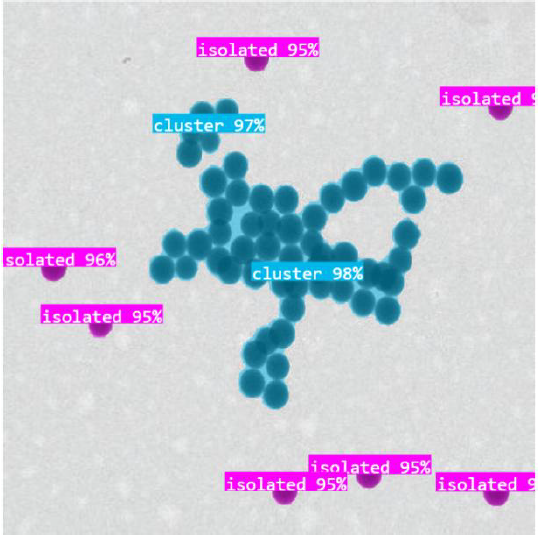}(f)}
	\end{minipage}
	\\
	\begin{minipage}[h]{0.3\linewidth}
		\center{\includegraphics[width=0.8\linewidth]{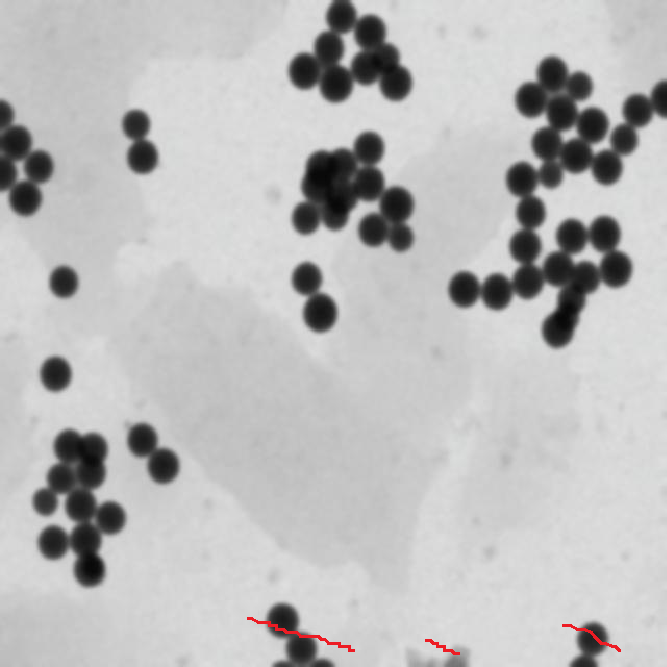}(g)}
	\end{minipage}
	\begin{minipage}[h]{0.3\linewidth}
		\center{\includegraphics[width=0.8\linewidth]{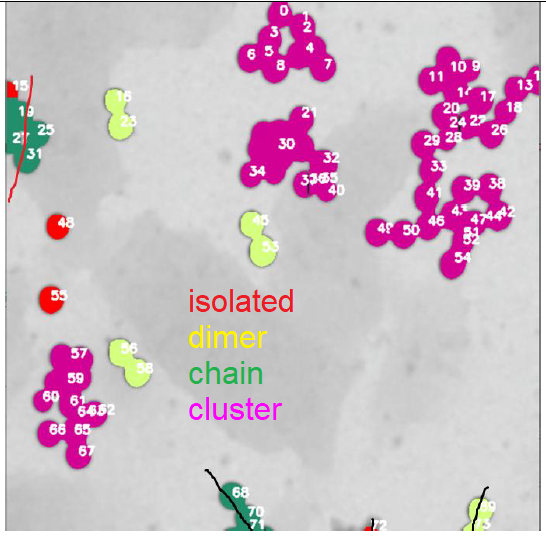}(h)}
	\end{minipage}
	\begin{minipage}[h]{0.3\linewidth}
		\center{\includegraphics[width=0.8\linewidth]{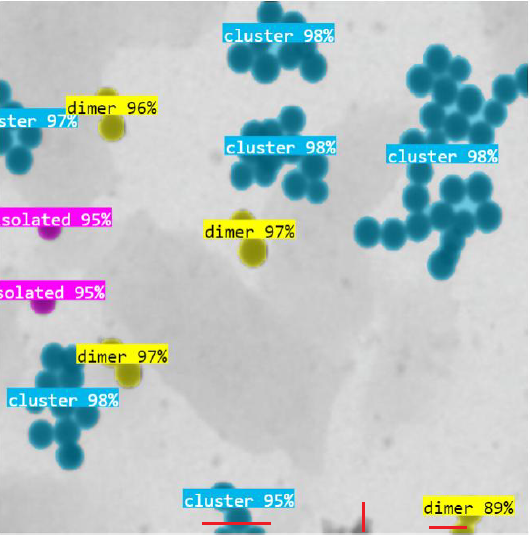}(i)}
	\end{minipage}
	\\
	\begin{minipage}[h]{0.3\linewidth}
		\center{\includegraphics[width=0.8\linewidth]{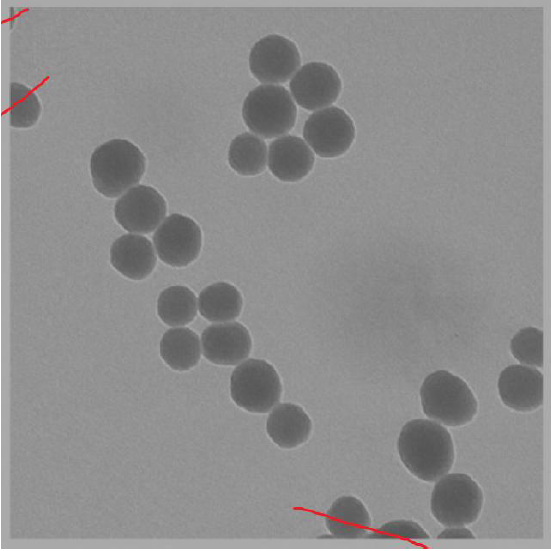}(j)}
	\end{minipage}
	\begin{minipage}[h]{0.3\linewidth}
		\center{\includegraphics[width=0.8\linewidth]{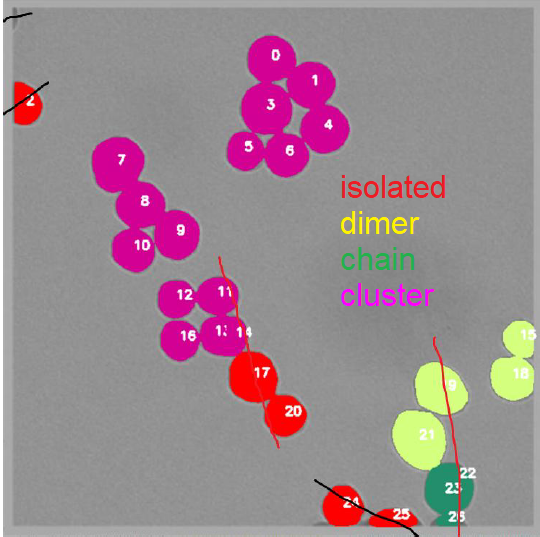}(k)}
	\end{minipage}
	\begin{minipage}[h]{0.3\linewidth}
		\center{\includegraphics[width=0.8\linewidth]{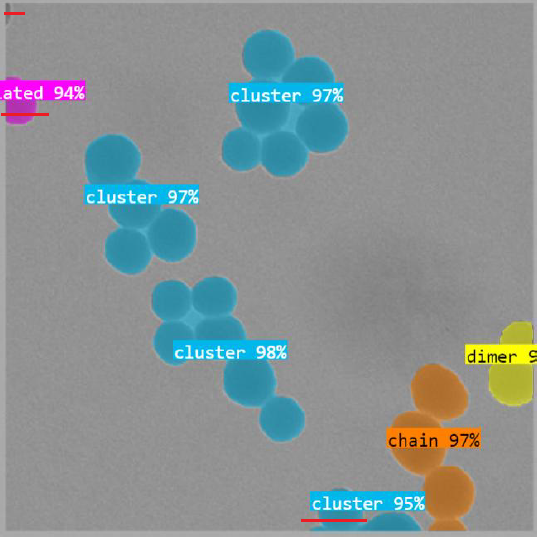}(l)}
	\end{minipage}
	\\
	\begin{minipage}[h]{0.3\linewidth}
		\center{\includegraphics[width=0.8\linewidth]{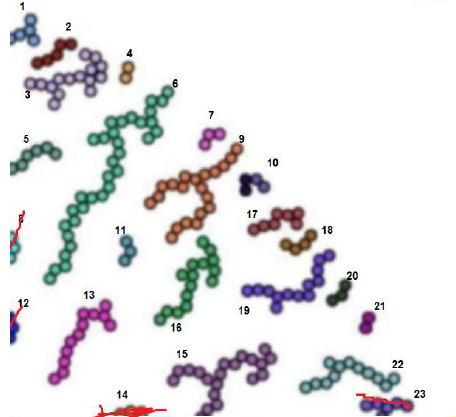}(m)}
	\end{minipage}
	\begin{minipage}[h]{0.3\linewidth}
		\center{\includegraphics[width=0.8\linewidth]{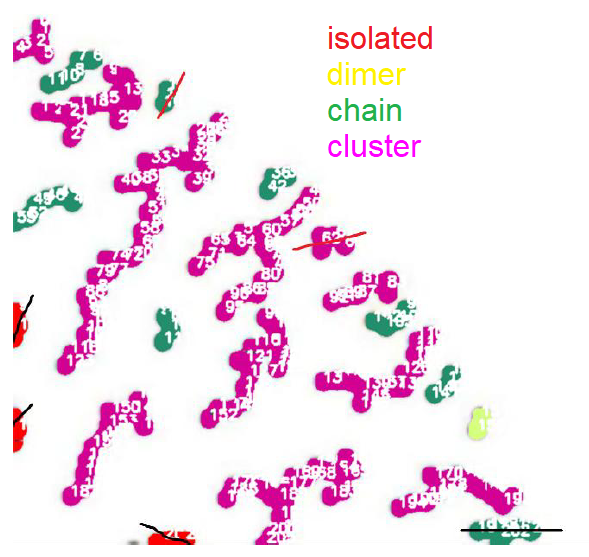}(n)}
	\end{minipage}
	\begin{minipage}[h]{0.3\linewidth}
		\center{\includegraphics[width=0.8\linewidth]{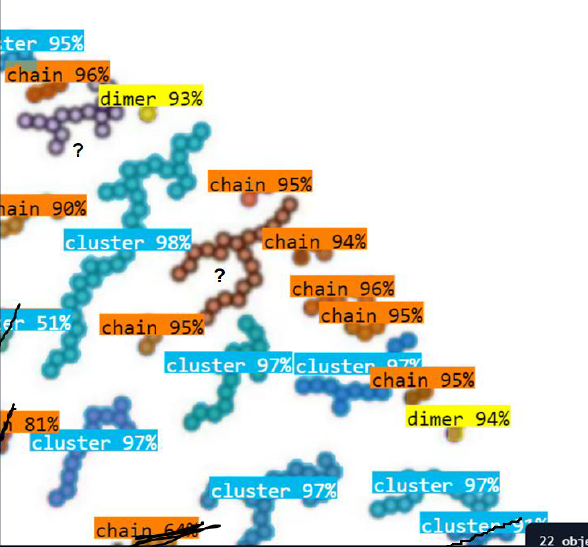}(o)}
	\end{minipage}
	\\
	\caption{Error analysis for two recognition methods: original (a) SEM, (d, g, j) TEM, and (m) simulation-based images; results of (b, e, h, k, n) thresholding and (c, f, i, l, o) neural network model. Image sources: (a) Alqedra et al., Licence CC BY 4.0~\cite{Alqedra2023}, (d, g) Abi-Ghaida et al., Licence CC BY 3.0~\cite{AbiGhaida2015}, (j) Yu et al., Licence CC BY~\cite{Yu2013}, and (m) generated image using the algorithm~\cite{Zolotarev2021}.}
	\label{fig:ErrorAnalysis}
\end{figure*}

We conducted an analysis of 50 images processed using the two aforementioned methods (these images were not used in the dataset for training the neural network model). For each image, we manually counted the number of particle assemblies $N_\mathrm{obj}$, excluding from the analysis those assemblies whose configuration type was ambiguous. Assemblies truncated by the edge of the image region often caused ambiguity. We also manually verified the processing results for each image, calculating the error using the formula $\delta = N_\mathrm{false}/ N_\mathrm{obj}$, where $N_\mathrm{false}$ is the number of objects from the original image that were falsely recognized or ignored. After processing all these images, we obtained the average error value $\delta_\mathrm{ave} = \left(\sum_{i=1}^{N_\mathrm{imgs}} \delta_i \right)/ N_\mathrm{imgs}$, where $N_\mathrm{imgs}$ is the number of images ($N_\mathrm{imgs} = 50$). A small example of such analysis is presented in Fig.~\ref{fig:ErrorAnalysis} and in Table~\ref{tab:ErrorAnalysis}. In the images, some particle assemblies are crossed out, as they were either excluded by us from consideration due to ambiguity or were falsely recognized. The assemblies that were not recognized at all (ignored by a program) are marked with question marks (Fig.~\ref{fig:ErrorAnalysis}).

\begin{table}[h]
	\centering
	\caption{Error analysis}
	\renewcommand{\arraystretch}{1.3} 
	\begin{tabular}{|p{0.4\linewidth}|p{0.1\linewidth}|p{0.1\linewidth}|p{0.1\linewidth}|}
		\hline
		Figure number & $N_\mathrm{obj}$ & $N_\mathrm{false}$ & $\delta$\\
		\hline
		Fig.~\ref{fig:ErrorAnalysis}b (thresholding) & 5 & 1 & 0.2 \\
		\hline
		Fig.~\ref{fig:ErrorAnalysis}c (ML) & 5 & 0 & 0 \\
		\hline
		Fig.~\ref{fig:ErrorAnalysis}e (thresholding) & 9 & 3 & 0.33 \\
		\hline
		Fig.~\ref{fig:ErrorAnalysis}f (ML) & 9 & 0 & 0 \\
		\hline
		Fig.~\ref{fig:ErrorAnalysis}h (thresholding) & 10 & 1 & 0.1 \\
		\hline
		Fig.~\ref{fig:ErrorAnalysis}i (ML) & 10 & 0 & 0 \\
		\hline
		Fig.~\ref{fig:ErrorAnalysis}k (thresholding) & 5 & 2 & 0.4 \\
		\hline
		Fig.~\ref{fig:ErrorAnalysis}l (ML) & 5 & 0 & 0 \\
		\hline
		Fig.~\ref{fig:ErrorAnalysis}n (thresholding) & 19 & 2 & 0.1 \\
		\hline
		Fig.~\ref{fig:ErrorAnalysis}o (ML) & 19 & 2 & 0.1 \\
		\hline
	\end{tabular}
	\label{tab:ErrorAnalysis}
\end{table}

\begin{figure}[htbp]
	\centering
	\includegraphics[width=0.99\linewidth]{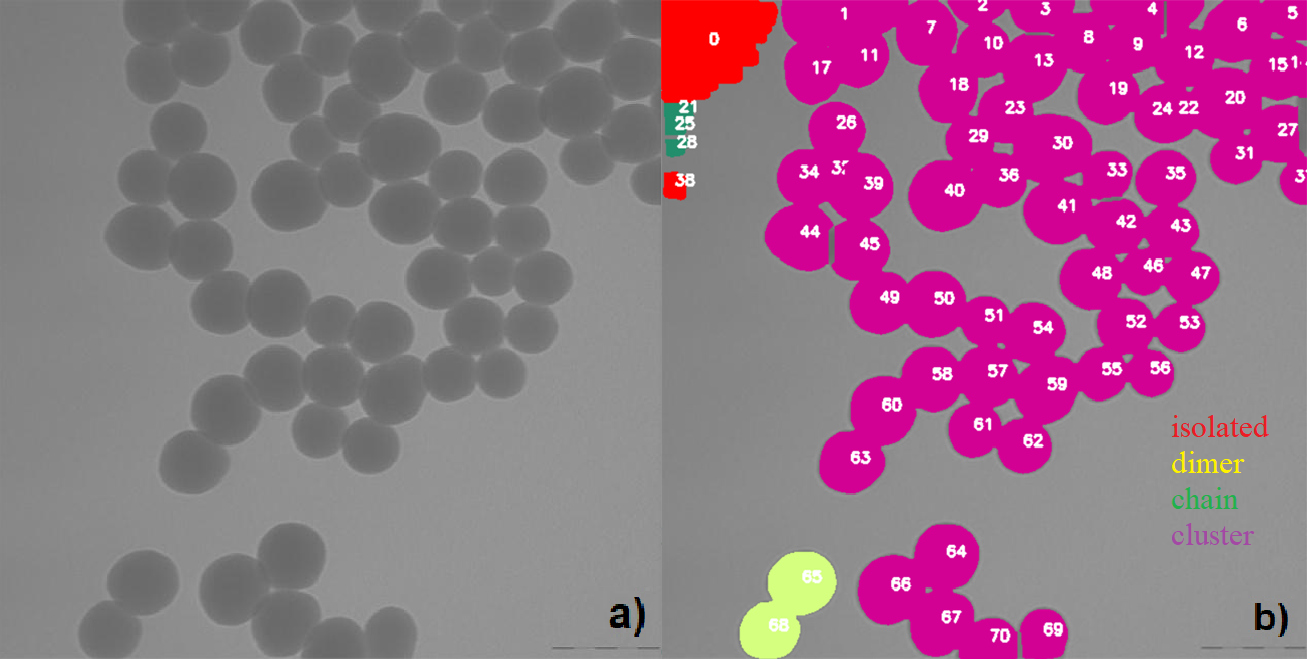}
	\caption{Image with noise: a) original TEM image (Zhao et al., Licence CC BY-NC 3.0) \cite{Zhao_2015}, b) analysis result.}
	\label{fig:ImageWithNoise}
\end{figure}

The study revealed that the identification error for colloidal assemblies using the thresholding method is approximately 33\%, while the error of the machine learning method is close to 3\%. This high error $\delta_\mathrm{ave}$ in the thresholding method is significantly influenced by inaccuracies in identifying individual particles within assemblies. The problem is related to the aforementioned irregularities and partial merging (where the boundaries of one particle are either erroneously included in another or, conversely, are truncated as the system interprets them as part of a neighboring object). Such errors arise due to particle overlap, insufficient image contrast, or background heterogeneity. All of this affects the distance between particle centroids, reducing the probability of correctly determining their neighborhood. The consequence is the incorrect identification of the assembly category. It is also worth noting that during the analysis of the thresholding method, noise on one image was identified as two isolated particles and one chain (Fig.~\ref{fig:ImageWithNoise}). In the image, it can be seen that the top-left corner is much darker than the rest of the image area. Therefore, it can be concluded that this was the cause of the false particle identification. We did not count this as an error since this case was unique (in the selected set of 50 images). This specific case further emphasizes the necessity of image preprocessing for this method. However, image preprocessing does not always guarantee accurate recognition of the formed assemblies. In some cases, image characteristics such as complex structure or high noise levels may persist even after processing, which limits the effectiveness of the thresholding method. In addition, there is another error here. Two particles are merged into one (particle number 1 in Fig.~\ref{fig:ImageWithNoise}b). Most likely, this is due to part of the area of these particles being cut off by the edge of the region.

Table~\ref{tab:MethodsComparison} presents a comparative analysis of the two methods. The key advantage of thresholding is its ability to identify individual particles within an assembly. However, in most aspects, the machine learning method demonstrates higher effectiveness. The runtime of the scripts for processing a single image for both methods was measured on an Intel Core i7-8700 CPU. For the YOLOv8 model, it was measured using already trained weights (the model training time is about three hours). Given that both methods complete in less than a second, the difference in time is insignificant.
\begin{table}[h]
	\centering
	\caption{Comparison of methods}
	\renewcommand{\arraystretch}{1.3} 
	\begin{tabularx}{\columnwidth}{|p{0.45\linewidth}|p{0.2\linewidth}|p{0.2\linewidth}|}
		\hline
		\textbf{Characteristic} & \textbf{Threshold processing} & \textbf{YOLOv8} \\
		\hline
		Determining the number of particles in an assembly & Yes & No \\
		\hline
		Ignoring labels & No & Yes \\
		\hline
		Takes noise for a particle & Rarely & Not detected \\
		\hline
		Accuracy & 67\% & 97\% \\
		\hline
		Preprocessing & Highly desirable & Rarely required \\
		\hline
		Operating time & 0.37 s & 0.73 s \\
		\hline
	\end{tabularx}
	\label{tab:MethodsComparison}
\end{table}

\section{Conclusion}

In this work, single-layer assemblies of colloidal particles (2D structures) were investigated with the aim of developing a method for identifying such assemblies based on image analysis. To solve this problem, two approaches were considered: image analysis using a modified thresholding method with the OpenCV library and a machine learning method using the YOLOv8 model.

A distinctive feature of the developed modified thresholding method is the automatic selection of the binarization type based on the circularity parameter calculated for each particle using the Watershed algorithm. The neighborhood and types of assemblies are determined from the coordinates of the centers of the detected particles. The image thresholding method, as applied in this study, demonstrated the necessity of preliminary manual image processing to achieve more accurate results. It was established that the presence of noise, artifacts, and other distortions in the original images leads to erroneous identification of some particles, which significantly affects the quality of the analysis. To mitigate this problem, preliminary cleaning of images from extraneous elements that can distort the data is required. Furthermore, the thresholding method showed an accuracy of approximately 67\%, which is directly related to the performance of the Watershed algorithm in identifying individual particles within assemblies. Despite its effectiveness, the Watershed algorithm has a certain margin of error in determining particle boundaries. This leads to inaccuracies in measuring the distances between them, which, in turn, affects the correctness of determining assembly categories. Of course, by manually tuning parameters for each specific image, even higher accuracy can be achieved. However, in the current work, we focused on full automation, which is crucial when processing a large number of images.

The best results were demonstrated by the machine learning method, specifically using the YOLOv8-Seg model for identifying colloidal assemblies (97\% accuracy). This approach showed high effectiveness, providing precise recognition of assemblies even in images with high noise levels. It can be hypothesized that the quality of particle identification depends on the surface concentration (a decrease in identification accuracy is observed in areas of high particle density). It is also worth mentioning that particles located at the image borders and partially cropped by the frame edge pose a challenge for correct cluster determination. Due to incomplete visual information (e.g., distorted geometry), the segmentation algorithm either fails to identify such objects as individual particles or misclassifies them into another configuration. This leads to identification inaccuracies. These observations provide a foundation for further research in this direction. Despite all this, the YOLOv8-Seg model has proven to be a powerful tool for automating the analysis process, which is particularly important when working with large datasets. Thus, the research results confirm the promise of using machine learning methods for colloidal assembly identification tasks. The software modules we developed and the trained model are integrated into ISANM and are freely available, which will subsequently enable researchers to develop more advanced algorithms and perform comparisons with our results (\href{https://isanm.space/}{https://isanm.space/})~\cite{Khusainova2025}.

\bmhead{Acknowledgements}

This work is supported by Grant No. 22-79-10216 from the Russian Science Foundation (\href{https://rscf.ru/en/project/22-79-10216/}{https://rscf.ru/en/project/22-79-10216/}).

\section*{Declarations}
\bmhead{Funding}

Grant number 22-79-10216 from the Russian Science Foundation (\href{https://rscf.ru/en/project/22-79-10216/}{https://rscf.ru/en/project/22-79-10216/}).

\bmhead{Conflict of interest}

The authors declare that they have no known competing financial interests or personal relationships that could have appeared to influence the work reported in this paper.

\bmhead{Data Availability}

The dataset used within the scope of the study will be shared upon reasonable request.

\bmhead{Author contribution}

LK: Programming, Dataset, Investigation,
Validation, Visualization, Writing~--- original draft. KK: Conceptualization,
Formal analysis, Funding acquisition, Methodology, Project
administration, Supervision, Writing~--- review and editing.

\bibliography{KhusainovaKolegov2025}
\end{document}